# Concept to assess the human perception of odour by estimating short-time peak concentrations from one-hour mean values. Reply to a comment by Müller et al.

Günther Schauberger [1,2,*], Martin Piringer [3], Rainer Schmitzer [1,4], Martin Kamp [1,5], Andreas Sowa [1,6], Roman Koch [1,7], Wilfried Eckhof [1,8], Ewald Grimm [1,9], Joachim Kypke [1,10], Eberhard Hartung [1,11]

[1] VDI Working Group „Emissions and their impact from livestock operations" (chair: E. Hartung)
[2] WG Environmental Health, University of Veterinary Medicine Vienna, Austria
[3] Central Institute of Meteorology and Geodynamics, Vienna, Austria
[4] Regierung von Mittelfranken, Ansbach, Germany
[5] Landwirtschaftskammer Nordrhein-Westfalen, Münster; Germany
[6] Steinfurt, Germany
[7] Landratsamt Fürstenfeldbruck, Fürstenfeld, Germany
[8] Ingenieurbüro Eckhof, Ahrensfelde, Germany
[9] Association for Technology and Structures in Agriculture (KTBL), Darmstadt, Germany
[10] LMS-Landwirtschaftsberatung, Schwerin, Germany
[11] Institut für landwirtschaftliche Verfahrenstechnik, Universität Kiel, Kiel, Germany

**Short Title**
Concept to assess the human perception of odour by the use of one-hour mean values

**Address**

* Corresponding author: Günther Schauberger
University of Veterinary Medicine
Department for Biomedical Sciences
Veterinärplatz 1
A 1210 Vienna
Austria

gunther.schauberger@vetmeduni.ac.at
++ 43 1 25077 4574

# Abstract

Biologically relevant exposure to environmental pollutants often shows a non-linear relationship. For their assessment, as a rule short term concentrations have to be determined instead of long term mean values. This is also the case for the perception of odour. Regulatory dispersion models like AUSTAL2000 calculate long term mean concentration values (one-hour), but provide no information on the fluctuation from this mean. The ratio between a short term mean value (relevant for odour perception) and the long term mean value (calculated by the dispersion model), called the peak-to-mean value, is usually used to describe these fluctuations. In general, this ratio can be defined in different ways. Müller et al. (2012), in a comment to Schauberger et al. (2012) which includes a statement that AUSTAL2000 uses a constant factor of 4, argue that AUSTAL2000 does not apply a peak-to-mean factor and does not calculate odour exceedance probabilities. Instead it calculates the frequency of so-called odour-hours by applying the relation between the 90-percentile of the instantaneous concentration and the hourly mean (Janicke and Janicke, 2007a), not between some peak value and the mean. According to Janicke and Janicke (2007a), the 90-percentile of the instantaneous concentration can in practice be estimated with sufficient accuracy from the hourly mean by using a factor of 4.

Having so far replied to Müller et al. (2012) we take additionally the opportunity to elaborate a little more on the peak-to-mean concept, especially pointing out that a constant factor independent of the stability of the atmosphere, the distance from and the geometry of the source, is not appropriate. On the contrary it shows a sophisticated structure which cannot be described by only one single value.

# 1  Introduction

Schauberger et al. (2012) present an empirical model to calculate the separation distance between livestock and residential areas to avoid nuisance by the emitted odour. The calculations to derive the parameters of the empirical model were done by the regulatory dispersion model AUSTAL2000. This dispersion model calculates one-hour mean values. Odour-hours are derived by multiplying the hourly mean value with a constant factor of 4 (Janicke and Janicke, 2007a; Janicke and Janicke, 2007b); see also the reply to Müller et al. (2012) above. If the calculated ambient odour concentration exceeds the threshold of 0.25 $ou_E/m^3$ (= odour detection threshold of 1 $ou_E/m^3$ divided by the constant factor 4) then this hour is counted as an odour-hour according to VDI 3940 Part 2 (2006). A more sophisticated model which would include the meandering of the plume was also discussed (Janicke and Janicke, 2007a), but not taken into account by the environmental protection agencies of the federal countries in Germany. It should be discussed if this model should be implemented instead to AUSTAL2000. Even if this constant factor 4 is not called "peak-to-mean factor" (see reply above), it serves the same purpose: to assess the odour perception on the basis of a one-hour mean value.

Schauberger et al. (2012) discuss the overestimation of separation distances by the dispersion model AUSTAL2000 in comparison with field measurements of odour for a low exceedance probability of 2% (irrelevance criterion) (Hartmann and Hölscher, 2007). Besides this constant factor 4, also other effects can in principle contribute to this overestimation.

(1) In general it is assumed that odorous substances behave like inert gases without chemical reactions or adsorption. There are some indications that this working hypothesis needs to be revised in some aspects. For cattle odour, for example, a change of the composition of odourous substances with travel time was found. Key odorants at the feedlot were volatile fatty acids and phenol compounds, but their relative importance diminished with downwind distance (Trabue et al., 2011).

(2) There is some evidence that odourous substances are present in particulate matter (PM) (Hoff et al., 1997; Liao and Singh, 1998). For pigs, Bulliner et al. (2006) and Cai et al. (2006) found VOCs and odourous substances in PM and their abundance was proportional to PM size. However, the majority of VOCs and characteristic pig odorant substances were preferentially bound to smaller-size PM. The reduction of PM and thus also odour substances can be calculated by a size-depending sedimentation velocity.

(3) Müller et al. (2012), in their point 2, state that also the meteorological boundary layer model implemented in AUSTAL 2000 might have an effect towards overestimation of small frequency values at large distances, discussed in detail by Janicke and Janicke (2011). The overestimation by AUSTAL2000 in comparison with field measurements was shown by Hartmann and Hölscher (2007).

In addition to the reply to Müller et al. (2012) already given above, we take the opportunity to elaborate more on the general problems of using dispersion models for the prediction of odour perception and the resulting annoyance and therefore discuss the assessment of short-term concentration values by the peak-to-mean concept in more detail.

## 2 Peak-to-mean concept

For the assessment of peak values, describing the biologically relevant exposure, often the so called peak-to-mean concept is used. The following overview should demonstrate that the peak-to-mean factor depends on several parameters, showing that the assumption of a constant peak-to-mean factor can only by used as a very rough estimate.

The step from the one-hour mean value (as output of the dispersion model) to an instantaneous odour concentration is shown in Fig. 1. For the one-hour mean value, the threshold for odour perception (here taken as 1 $ou_E/m^3$) is not exceeded. Taking mean values over 10 minutes, one concentration value exceeds the threshold. For the short term mean values of 12 s, concentrations in the range of 5 to 6 $ou_E/m^3$ can be expected, which means a distinct odour perception over several breaths. Fig. 1 shows that the shorter the selected time interval, the higher the maximum concentration. For the shortest period of 12 s, a new feature of the time series can be seen. Besides 12 s intervals with odour concentrations above zero, a certain percentage of zero observations can be expected. The frequency of non-zero intervals is called intermittency $i$.

Given a mean concentration over one hour, the mean value of a shorter period can be calculated using the well known relationship (cit. by Smith (1973)):

$$\frac{C_p}{C_m} = \left(\frac{t_m}{t_p}\right)^u \quad (1)$$

with the mean concentration, $C_m$, calculated for an integration time of $t_m$ and the peak concentration $C_p$, for an integration time of $t_p$. Hinds (1969) showed the evidence of this relationship by measurements.

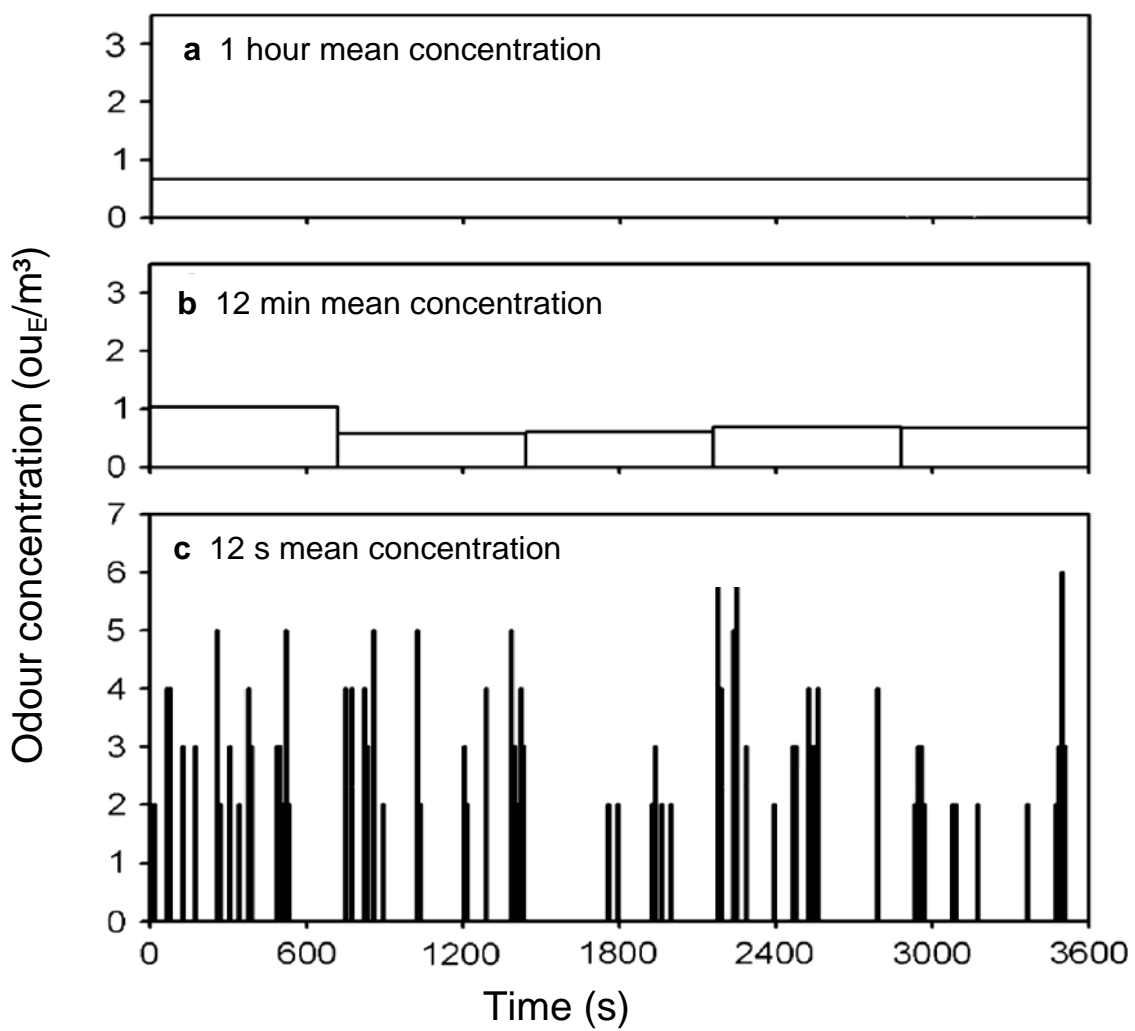

Fig. 1    Time course of the odour concentration ($ou_E/m^3$) for thress time intervals. (a) one-hour mean value (e.g. output of the dispersion model AUSTAL2000), (b) 12-min and (c) 12-s mean odour concentrations observed at a single receptor point during a field study. The 12-s mean values were recorded and subsequently used to calculate 12-min and one-hour mean concentrations (modified from Nicell (2009)).

According to the relationship above, the peak-to-mean factor is defined by $F = C_p/C_m$. The open question is the definition of the peak value $C_P$.

This peak value $C_p$ can be defined manifold (Gross, 2001; Klein and Young, 2010). The following definitions are used frequently: (1) $C_p = C_m + \sqrt{\sigma}$, i.e. the peak value is defined by the mean value and the standard deviation. The quotient between the standard deviation $\sqrt{\sigma}$ and the mean value $C_m$ is called fluctuation intensity $i = \sqrt{\sigma}/C_m$ therefore the peak-to-mean factor on the basis of the fluctuation intensity is $F_i = i + 1$. (2) The peak value is defined by the 90-percentile, so $F_{90} = C(p = 0.90)/C_m$, (3) the peak value is defined the 98-percentile or 99-percentile (Klein and Young, 2010) or (4) by the maximum $F_{max} = C_{max}/C_m$ (Klein and Young, 2010).

Especially for Germany, the peak value $C_P$ is well defined by the comparison between empirical field measurements (VDI 3940 Part 2, 2006) and dispersion model calculations. If we assume that the assessor sniffs every 10 seconds to decide if the sample smells, then we get 360 breaths (sample size) during one hour. In the German jurisdiction an hour is counted as a so called odour-hour if at least 10% of the 360 breathes can be evaluated as odourous. For practical reasons (VDI 3940 Part 2, 2006), only a period of 10 minutes (60 breaths) is used as a sample to judge a certain hour. If 6 out of 60 periods (10 minutes) are assessed as odours by a panellist, this defines an odour-hour. Therefore the 90-percentile is used to define the peak value $C_P$ with $t_p$ = 1 s to assess the incidence of an odour-hour. All hour values which lie above this criterion are called odour-hour, and the exceedance probability is then called frequency of odour-hours.

The influence of the integration interval $t_p$ of the peak concentration $C_p$ is shown schematically in Fig. 2. The shorter the integration interval, the higher the variance of the ambient concentration $C$. On the other hand, the graph shows the sensitivity of the definition of the peak value, which is defined by a certain percentile of the cumulative distribution function CDF. In this example the peak-to mean factor $F = C_p/C_m$ varies for the two integration intervals $t_{p1}$ and $t_{p2}$ between $F_i$=1.35, $F_{90}$=1.47 and $F_{98}$=1.85 and $F_i$=2.23, $F_{90}$=2.79 and $F_{98}$=5.17, respectively. This example shows the importance of a proper definition of the determination of the peak value.

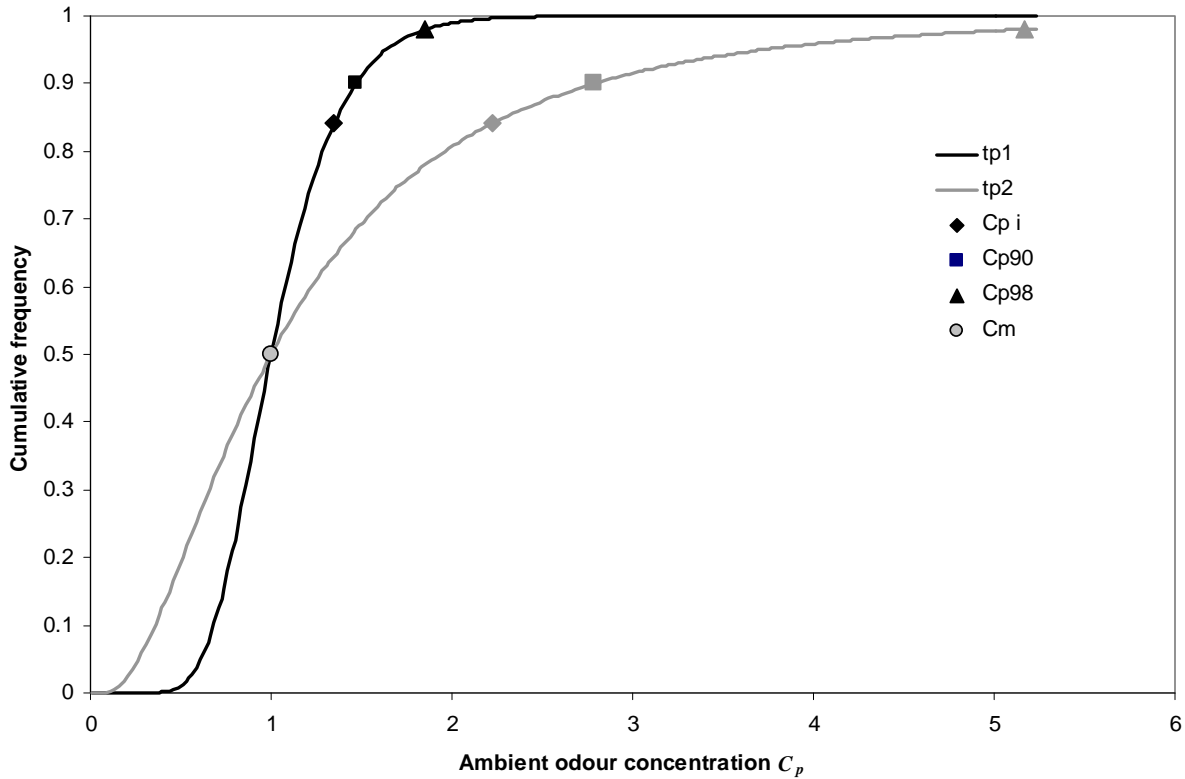

Fig. 2  Schematic diagramm showing the cumulative distribution function (log-normal) for two different integration intervalls $t_p$ with $t_{p1}$ > $t_{p2}$, of the peak concentration $C_p$. The mean value (one hour mean) $C_m$=1 ou$_E$/m³ is the same for both CDFs. The peak values are defined by the fluctuation intensity $i$ ($C_{pi}$), the 90-percentile ($C_{p90}$) and the 98-percentile ($C_{p98}$), showing the influence of integration interval $t_p$ and the determination of the peak value $C_p$ due to a certain percentile. The peak-to mean factors $F = C_p / C_m$ vary for the two integration intervals $t_{p1}$ and $t_{p2}$ between $F_i$=1.35, $F_{90}$=1.47 and $F_{98}$=1.85 and $F_i$=2.23, $F_{90}$=2.79 and $F_{98}$=5.17, respectively.

The assessment of maximum values for shorter periods than one hour is not only relevant for environmental odour but also for toxic and inflammable pollutants (Hanna, 1984b; Hilderman et al., 1999; Mylne, 1988). This estimation by one-hour mean values can lead to an underestimation of the impact of the ambient concentration. This error depends on the observed impact of the ambient concentration. In many cases the health impact is described by a non-linear dose response function (Hilderman and Wilson, 1999). Especially health related phenomena show such a relationship with the ambient concentration $C$ which can be described by a power function $C^\alpha$ with an exponent α (Miller et al., 2000) in the range between 1.0 and 3.5. Some chemicals show an exponent between 2.0 and 3.0 for the toxicity and fatalities. Only if the exponent $\alpha = 1$ then the concentration can be determined by a mean value. However, if $\alpha > 1$, then the use of the mean concentration will underestimate the impact of the substances. The health effects of toxic gases in this context are described in detail by Hildermann (1997).

# 3  Parameters which influence the concentration fluctuation (peak-to-to mean factor)

The assessment of a concentration value for a shorter integration time on the basis of a one-hour mean can be calculated by a peak-to-man factor. This conversion depends on the dilution process in the atmosphere which is predominantly influenced by turbulent mixing. The following predictors are discussed, which influence the concentration fluctuation (Hanna and Insley, 1989; Olesen et al., 2005):

1. Stability of the atmosphere
2. Intermittency
3. Travel time or distance from the source
4. Lateral distance from the axis of the wake
5. Geometry of the source (emission height and source configuration)

Turbulent mixing in the atmosphere depends strongly on the stability of the atmosphere. The stability can be determined e.g. by discrete stability classes or by the Monin–Obukhov length. The influence of the stability of the atmosphere on the peak-to-mean value is calculated e.g. by equation (1), using $t_m = 3600$ s (calculated one-hour mean) and $t_p = 5$ s (duration of a single breath). The peak-to-mean factors, depending on atmospheric stability, are derived by the exponent $u$ of Smith (1973) and Trinity Consultants (1976) (cit. by Beychock (1994) (Table 1). Lung et al. (2002) found, from measurements ($t_p = 1$ s, $C_p$ defined to be the maximum) in the near field of the source, a peak-to-mean factor in the range of $4 < F_{max} < 99$, Santos et al. (2009) could show the influence of the stability of the atmosphere on the exponent $u$ of equation (1).

Table 1  Maximum peak-to-mean factor $F=C_p/C_m$ calculated by equation (1) for an integration time for the peak value $t_p = 5$ s and the mean value $t_m = 3600$ s, depending on atmospheric stability by using values of Smith (1973) and Trinity Consultants (1976) (cit. by Beychock (1994)

| Stability class | Smith (1973) | | Trinity Consultants (1976) | |
| --- | --- | --- | --- | --- |
| | Exponent $u$ | $C_p/C_m$ | Exponent $u$ | $C_p/C_m$ |
| unstable | 0.64 | 67.4 | 0.68 | 87.7 |
| slightly unstable | 0.51 | 28.7 | 0.55 | 37.3 |
| neutral | 0.38 | 12.2 | 0.43 | 16.9 |
| slightly stable | 0.25 | 5.2 | 0.30 | 7.2 |
| stable | 0 | 1.0 | 0.18 | 3.3 |
| very stable | 0 | 1.0 | 0.18 | 3.3 |

As a relationship between the fluctuation intensity $i$ and the peak-to-mean factor $F$ Lung et al. (2002) found

$$F = 1 + \alpha\, i^2 \qquad (2)$$

with $\alpha = 3.6$ and the fluctuation intensity $i = \sqrt{\sigma}/C_m$ (Fig. 3). Here, the peak-to-mean factor is related to a peak concentration measured for an integration time of 1 s and the one-hour mean. Various peak-to mean factors $F$ (according to the definition of the peak concentration by the standard deviation, the 98-percentile or the maximum) were measured by (Klein and Young, 2010), showing the increase of the factor $F$ with the selected percentile.

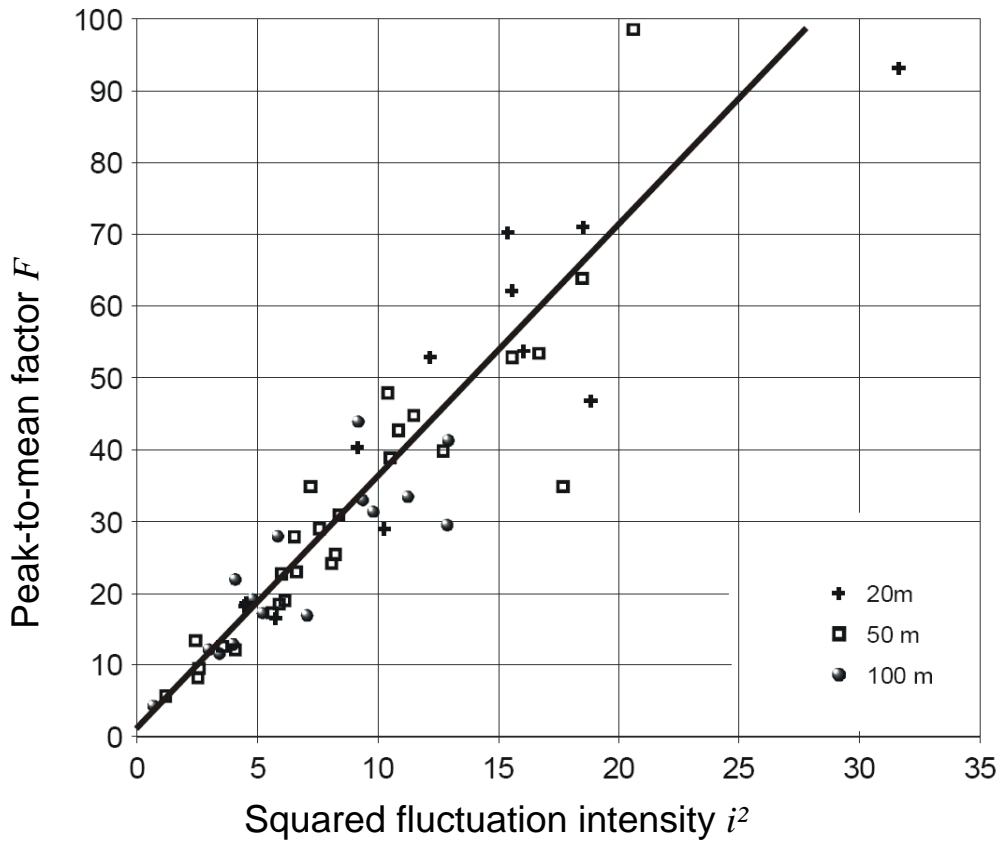

Fig. 3  Relationship between the squared fluctuation intensity $i^2$ and the peak-to-mean factor $F$ measured in the vicinity of a point source (distance beteen 20 and 100 m) (from Lung et al. (2002))

The fluctuation intensity $i$ is also related to the intermittency $\gamma$ by

$$i^2 = \frac{\beta}{\gamma} - 1 \qquad (3)$$

The parameter $\beta$ is determined to lie between $\beta = 2$ (Hanna, 1984a), $\beta = 3$ (Best et al., 2001), and $\beta = 3.6$ (Lung et al., 2002). Other functions describing this relationship can be found by (Klein and Young, 2010).

The lateral distance $y$, normalised by the lateral dispersion parameter $\sigma_y$ from the plume centre line shows a strong influence on the fluctuation intensity (Best et al., 2001; Hinds, 1969; Katestone Scientific, 1998; Løfstrøm et al., 1996). The following function describes this relationship:

$$i(x, y) = i(x) \exp\left(\frac{y^2}{a\,\sigma_y^{\,2}}\right) \qquad (4)$$

with $a = 2$ suggested by Best et al. (2001) and $a = 4$ by Katestone Scientific (1998).

The distance dependent reduction of the fluctuation intensity can be calculated by a model published by Best et al. (2001) and Katestone Scientific (1998) as a function of atmospheric stability and the geometry of the emitting source. Løfstrøm et al. (1996) and Hanna (1984b) suggested in each case a model for the fluctuation intensity $i$ as a function of the distance $x$ and the lateral distance $y$. These models are based on the dispersion parameters $\sigma_y$ and $\sigma_z$, which are known from the Gaussian dispersion model. The previous two models and the following model (Schauberger et al., 2000) for the decrease of the peak-to-mean factor with distance from the source can be used as a post-processing tool for dispersion calculations.

$$F = 1 + (F_o - 1)\exp\left(-0.7317\frac{T}{t_L}\right) \qquad (5)$$

where the peak-to-mean ratio of Table 1 is used as $F_0$ which is modified by an exponential attenuation function of $T/t_L$, where $T = x/v$ is the time of travel with the distance $x$, and the mean wind velocity $v$, and $t_L$ is a measure of the Lagrangian time scale (Mylne, 1992), which involves knowledge of the standard deviations of the three wind components. The calculation of the Lagrangian time scale in this context was improved by Piringer et al. (2007).

The reduction of the fluctuation intensity with the distance is shown exemplarily by empirical data provided by Mylne (1990) (Fig. 4). The fluctuation intensity seems to approach a constant value of 1.0 ± 0.3 (Mylne, 1990).

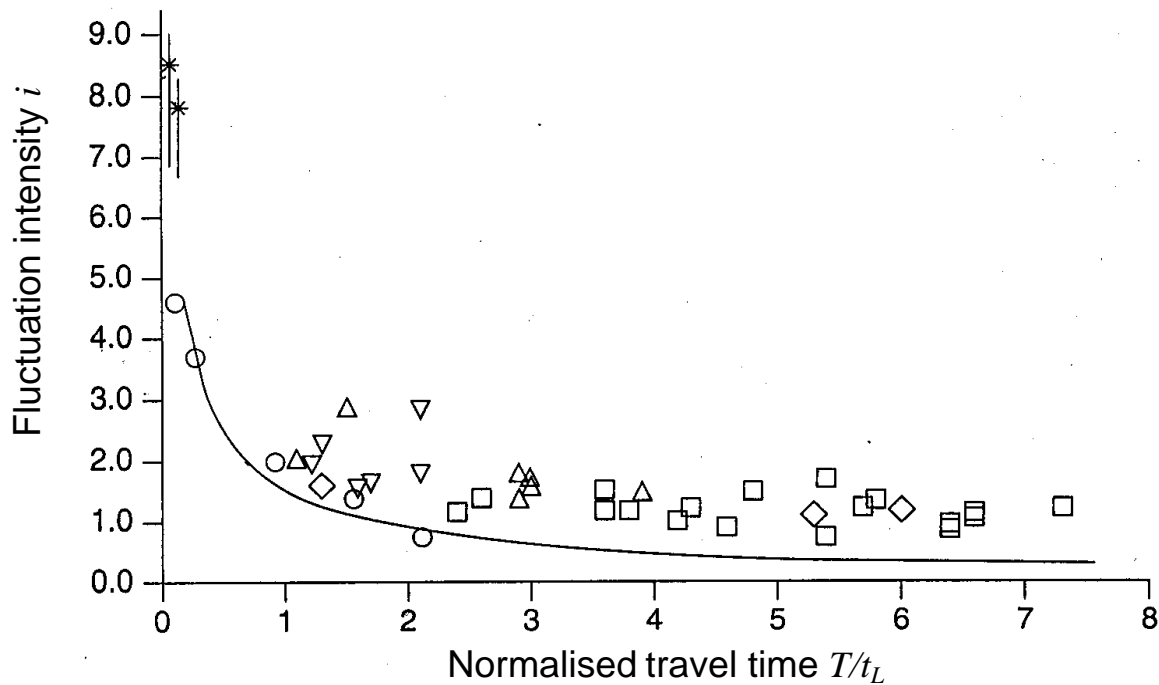

Fig. 4  Relationship between the fluctuation intensity $i$ and the normalised travel time $T/t_L$ (from Mylne (1990))

The influence of the geometry of the source (elevated point source vs. area source) is discussed in detail by Katestone Scientific (1998), showing lower peak-to-mean rations for area sources compared to elevated point sources. Best et al. (2001) suggested an exponent $u = 6/17$ and $u = 3/14$ for point and line sources, respectively. Fackrell and Robins (1982) and Mylne (1993) found higher fluctuations in elevated sources. For area sources the reduction of the fluctuation intensity with distance was lower compared to elevated sources.

Nevertheless, also constant peak-to-mean factors are in use. Some examples: peak-to-mean factor $F = 10$ for the pervious regulatory Gaussian dispersion model in Germany (Rühling and Lohmeyer, 1998), the Danish model with $F = 7.8$ (Olesen et al., 2005), and the constant factor $F = 4$ for AUSTAL2000.

# 4  Conclusions

This reply is inspired by a comment by Müller et al. (2012) to Schauberger et al. (2012). The latter includes a statement that AUSTAL2000 uses a constant peak-to-mean factor of 4. Müller et al. (2012) argue that AUSTAL2000 does not apply a peak-to-mean factor and does not calculate odour exceedance probabilities. Instead it calculates the frequency of so-called odour-hours by applying the relation between the 90-percentile of the instantaneous concentration and the hourly mean. According to Janicke and Janicke (2007a), the 90-percentile of the instantaneous concentration can in practice be estimated with sufficient accuracy from the hourly mean by using a factor of 4. Even if this constant factor 4 is not called "peak-to-mean factor", it serves the same purpose: to assess the odour perception on the basis of a one-hour mean value.

In presenting the well-known concept of the peak-to-mean value to assess a short-time concentration in some detail, it is shown that this is an important tool to estimate biologically relevant exposure (e.g toxicity, odour perception). Concentration fluctuations in dispersing plumes have been investigated experimentally, both in wind-tunnel studies and in field dispersion trials, over many years. Many authors could show the influence of various predictors on the peak-to-mean ratio: distance from the source, stability of the atmosphere, lateral distance from the centre of the plume, the geometry of the source (area, volume of point source), and height of the receptor point. Therefore we are convinced that a constant factor for the ratio between the 90-percentile of the instantaneous concentration (mean over about one second) and the hourly mean, which is used for the regulatory dispersion model AUSTAL2000 in Germany, does not fulfil these requirements, especially to assess the 2-%-exceedance probability which is used for the irrelevance criterion. On the other hand, we see the necessity to define the peak-to-mean value in a reliable manner to omit misunderstandings. Talking about a peak-to-mean factor, it is often not clear how the peak value is defined. We suggest to use the well defined method by VDI 3940 Part 2 (2006) based on field measurements with sniffing teams, which defines the peak value via the 90-percentile of short-term (1 s) concentration values. Detailed investigations, comparing field measurements with model calculations, would be helpful to improve the prediction of odour perception by dispersion models.